# EnergAIze: Multi Agent Deep Deterministic Policy Gradient for Vehicle-to-Grid Energy Management

Tiago Fonseca[1,2]*, Luis Ferreira[1,2]*, Bernardo Cabral[1,2], Ricardo Severino[1,2], Isabel Praça[2,3]
INESC-TEC[1], Polytechnic of Porto - School of Engineering[2], GECAD[3]
Porto, Portugal
{calof*, llf*, bemac, sev, icp}@isep.ipp.pt

*Abstract*—The rising adoption rates and integration of Renewable Energy Sources (RES) and Electric Vehicles (EVs) into the energy grid introduces complex challenges, including the need to balance supply and demand and smooth peak consumptions. Addressing these challenges requires innovative solutions such as Demand Response (DR), Renewable Energy Communities (RECs), and more specifically for EVs, Vehicle-to-Grid (V2G). However, existing V2G approaches often fall short in real-world applicability, adaptability, and user engagement. To bridge this gap, this paper proposes EnergAIze, a Multi-Agent Reinforcement Learning (MARL) energy management algorithm leveraging the Multi-Agent Deep Deterministic Policy Gradient (MADDPG) algorithm. EnergAIze enables user-centric multi-objective energy management by allowing each prosumer to select from a range of personal management objectives, thus encouraging engagement. Additionally, it architects' data protection and ownership through decentralized deployment, where each prosumer can situate an energy management node directly at their own dwelling. The local node not only manages local EVs and other energy assets but also fosters REC wide optimization. EnergAIze is evaluated through case studies employing the CityLearn simulation framework. The results show reduction in peak loads, ramping, carbon emissions, and electricity costs at the REC level while optimizing for individual prosumers objectives.

## I. INTRODUCTION

Renewable Energy Sources (RES) and Electric Vehicles (EVs) are emerging as pivotal players in the shift towards a low-carbon smart grid [1]. Small-scale wind and photovoltaic (PV) solar production are fostering a shift in how energy is produced, from a centralized infrastructure to a distributed generation where individuals directly engage with the energy grid. Such individuals are designated as prosumers, reflecting their dual role as both producers and consumers [2].

### A. Integration Challenges

Despite stated benefits, the surge in the adoption levels of EVs and RES, although promising for environmental sustainability, can also pose a series of infrastructural, control, technological, and societal challenges with negative impacts on the grid [3]. According to [4], at EV adoption levels of 25%, peak load can grow by 30% due to the EVs exerting pressure on the grid for example when arriving home after work, which is also the time the PV panel energy production decreases due to the sunset.

The environmental potential of EVs can only be maximized if the electricity they use originates from clean, renewable sources. As such, the stated integration challenges not only negate the advantages of RES and EVs but can also result in overall costlier electricity [5]. Therefore, researchers point to the need for a significant transformation in how we generate, store, distribute, and consume energy [6], towards intelligent energy management strategies.

### B. Energy Management

Intelligent energy management provides solutions to the mentioned problems by controlling, shifting, optimizing, and scheduling the use of energy resources more efficiently [7]. Central to these solutions are Demand Response (DR), and Renewable Energy Community (REC) management [8], which strive to balance energy supply and demand, reducing costs, and enhancing grid stability [9]. Both depend on prosumer's energy flexibility, which can be represented as Flex Offers (FOs) [10], and in essence are the prosumer's availability to delay or advance their consumptions in time.

Moreover, the Vehicle-to-Grid (V2G) concept leverages the flexibility of EVs for power grid management, enhancing load shifting capabilities. [11]. In V2G, EVs are turned into energy storage units that not only draw power from the grid but can also feed electricity back into the grid during periods of high energy demand. For instance, EVs can be charged when solar panels are at peak output, then either use this renewable energy to commute or send it back to the grid during times of high demand or low renewable generation. This balances the grid, cuts reliance on peak power sources like gas, and offers EV owners a return on their investment by selling excess energy. In parallel, simpler strategies, such as smart charging Grid-to-Vehicle (G2V) can optimize energy consumption during high demand periods [12].

### C. Gaps and Contributions

The coordinated and autonomous management of flexible assets into the energy grid, and specifically EVs into intelligently managed REC is still in its infancy [13]. The prevailing strategies in the real-world often utilize Time-of-Use (ToU) pricing to incentivize EV owners to plug in their cars during cheaper periods. However, such strategies require prosumers to actively manage and be aware of their consumption and energy prices, which can be tedious and time-consuming, and lead to prosumer disengagement [14].

On this notice, researchers are turning into intelligent management approaches, hereby called Energy Management Systems (EMS), that use advanced algorithms to optimize EV charging given factors like grid conditions, energy prices, vehicle usage patterns, prosumer flexibility, among others [15]. Applied algorithms range from Machine Learning (ML), to Model Predictive Control (MPC), meta-heuristics and extant optimal control algorithms (Section II). However, these technologies are not yet ready for real-world deployment due to: limitations in scalability, no integration of V2G within broader REC optimization, concerns over data ownership, and the need for further refinement in adaptive learning to cater to prosumers objectives, motivating their participation.

Given the identified research gap, this paper introduces EnergAIze, a Multi Agent Reinforcement Learning (MARL) EMS framework designed for managing the energy flexibility of EVs, and other energy flexible assets, such as PVs, and Heat Pumps within RECs. Contributions can be resumed as:

- EnergAIze applies the Multi-Agent Deep Deterministic Policy Gradient (MADDPG) algorithm, emphasizing user-centric energy management and allow prosumers to select one of the following personal objectives for optimization: i) cost minimization, ii) self-consumption maximization, iii) carbon emission reduction.
- EnergAIze requires minimal input from the prosumer, for performing V2G with only the departure time and the required State of Charge (SoC) at departure for the EV.
- EnergAIze encourages a REC-oriented approach where prosumers not only optimize their energy but also support the REC by trading energy with neighbours.
- Decentralized architecture design, powered by edge computing, which ensures privacy and data sovereignty for prosumers engaging in this strategy.
- Demonstration EnergAIze's effectiveness with a simulation of V2G within a REC scenario.

Section II explores related projects in this domain. Section III describes the applied algorithm. The simulation is described in Section IV. Section V showcases results, while Section VI concludes and outlines future directions.

## II. RELATED WORK

Recent research has highlighted the critical role of optimizing EV charging schedules within the energy grid. Traditional optimization methods struggle to manage these complexities, propelling interest towards RL techniques with its ability to adapt and learn from complex systems [16]. Deep RL (DRL) techniques have been applied to maximize renewable energy consumption and the SoC of EVs at departure. Further research has combined RL with Time of Use (ToU) energy pricing to minimize infrastructure costs, while Deep Q-Network-based RL (DQN-RL) algorithms have been tested against real-world data to validate their charging strategies effectiveness [17]. However, these approaches have led to some data security and scalability concerns [18]. Moreover, Multi-Agent Reinforcement Learning (MARL), as it allows for multiple decentralized agents to learn cooperatively, offering scalable and robust solutions for complex energy network management [19]. For example, MARLISA, introduced in [20] demonstrates the potential of MARL for decentralized load shaping in energy systems, achieving superior performance reducing peak loads. MERLIN in [21] tackles RL's challenges by leveraging independent battery control policies and transfer learning.

Despite these advancements, to the best of the authors knowledge, there is limited exploration into integrating V2G with the diversity of control of other REC assets. As pointed by the work in [5], the best strategy for energy management should consider the diversity and heterogeneity of multiple assets, gathering their energy flexibility for most effective management. Additionally, there is a lack of research on multi-objective optimization that enables each prosumer to select their goals within the broader optimization scope.

## III. ENERGAIZE SYSTEM ARCHITECTURE

### A. Multi-Agent Formulation for Energy Management

At the basis of any MARL formulation is the Markov Game [22]. The Markov Game extends single agent RL Markov Decision Process and considers multiple agents, in this case each referring to the EMS algorithm managing a dwelling's energy. These agents operate within discrete time steps represented by t, with t∈{1,2,...,N+}.

### 1) Core Components

The core components for the MARL approach are depicted in Figure 1 and a brief explanation follows:

- **Agent** $i$ for $i \in \{1,2,...,N\}$, represents one of $N$ EnergAIze decision-making nodes deployed at each REC dwelling.
- **Observation,** denoted as $O^i_t$, reflects the localized perspective of agent $i$ at a given time $t$. $O^i_t$ within a dwelling can be captured, for example, by real-world IoT sensor readings. In the implementation, $O^i_t$ can be expressed, but not limited to, $O^i_t = \{O^{NSL}_t, O^{SG}_t, O^{CN\_SoC}_t, O^{CN\_EDT}_t, O^{CN\_SoC\_D}_t ...\}$. Each element within this set is an individual input within a dwelling. $O^{NSL}_t$ represents the Non-Shiftable Load in kWh within dwelling $i$ at time $t$; $O^{SG}_t$ denotes the PV Solar Generation in kWh within dwelling $i$, $O^{CN\_SoC}_t$ indicates the State of Charge (SoC) of an EV connected to a Charger $CN$ within dwelling $i$; $O^{CN\_EDT}_t$ states the Estimated Departure Time and $O^{CN\_SoC\_D}_t$ the required SoC at Departure for an EV. Observations on the forecast of PV production or on the arrival time of an EV can also be provided. Besides local observations, $O^i_t$ it includes Energy Prices $O^{EP}_t$, Carbon Emissions $O^{CE}_t$ observations. One-hot encoding is used for categorical observations such as Day of the Week.
- **State,** with $S_t=[O^1_t, O^2_t,...,O^N_t]$, is the aggregated global view of the environment, formed by pooling together the observations from all dwellings $i$ at time $t$.
- **Action,** denoted $A^i_t$ reflects the agent $i$ decided actions A at time $t$. $A^i_t$ can be expressed as $A^i_t=\{A^{HS}_t, A^{N\_EV}_t,...\}$, where each element is an individual action within a dwelling. For instance, $A^{HS}_t$ represents the action (turn on or off) applied to the Heating Storage (electric heater) management, and $A^{CN\_EV}_t$ denotes the action to charge or discharge a connected EV by some amount of energy.
- **Environment,** denoted as ε, this concept represents the REC as a whole. ε is where Agents $i$ operate, $O^i_t$, are collected, and $A^i_t$ are applied. The real-world EnergAIze energy management environment can be defined as: dynamic, continuously evolving in response to unpredictable factors like weather and energy demand; partly inaccessible; continuous, due to the range of possible action values for charging; non-deterministic, as outcomes of actions cannot be predicted with certainty due to external influences.

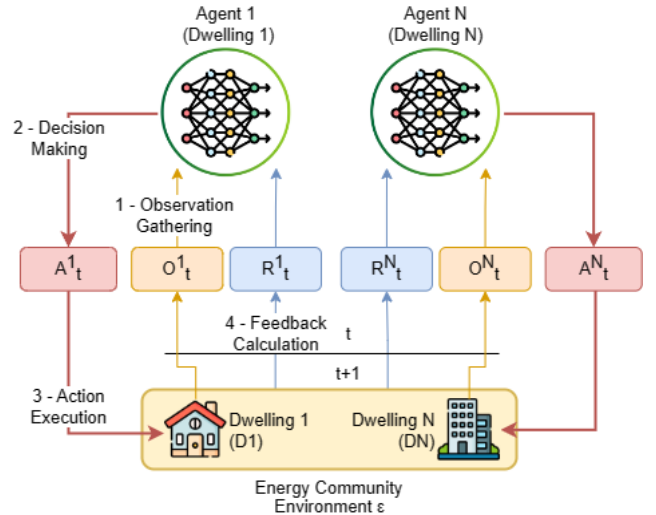

*Figure 1 - Problem contextualized within MARL*

*2) Functional Aspects and Feedback Components*

In MARL systems, two elements govern how agents act and learn, the **Policy Function** $\pi^i(A^i_t|O^i_t)$ determines the $A^i_t$ that each agent $i$ will take based on its current observation $O^i_t$. This is the function that is learned and refined during training and guides the dwelling's energy management at deployment. The **Reward** ($R^i_t$) is the feedback that agent $i$ receives from the environment after executing actions $A^i_t$. (Section III.D.). The reward will be composed of local and REC wide optimization evaluation, in a multi-level approach as defined by [23].

*3) Logical Flow*

Fig. 1. depicts the MARL process that ties the presented components together.

1. **Observation Gathering:** At the start of each time step $t$, each agent $i$ gathers local observations $O^i_t$ about its dwelling environment, which can be captured by real-world IoT sensor readings or simulated.
2. **Decision Making:** Based on $O^i_t$, the agent consults its policy function $\pi^i(A^i_t|O^i_t)$ to decide on the values for $A^i_t$.
3. **Action Execution:** Once decided, the agent $i$ applies $A^i_t$ in the environment ε, evolving to the next time step by using the State Transition Function $f(O^i_{t+1} | O^i_t, A^i_t)$.
4. **Feedback Calculation:** Following action execution, each agent $i$ garners feedback from the environment in the form of a reward $R^i_t$. This reward gauges how well the actions $A^i_t$. helped on reaching the agent's objectives and the overarching goals of the system.

Note on the decentralized nature of the steps described, emphasizing how each agent within a MARL formulation acts autonomously relying solely on its local observations to decide its actions, discarding the need to rely on communication with other agents during runtime.

*B. Physical Deployment*

Fig. 2. illustrates the architected decentralized deployment of EnergAIze into an REC with a set of dwellings, each with a different set of assets. EnergAIze is architected to operate on a decentralized model approach, empowering individual dwellings and prosumers to autonomously optimize energy flexibility within REC objectives. This decentralization is made possible through an edge computing framework, such as the one present in [24]. In this approach, every dwelling has an edge computing device.

The edge computer holds the agent $i$ which, through its trained policy function, reasons on $O^i_t$ collected from its dwelling and produces $A^i_t$ to guide the energy consumption. Observations come from the local sensors and smart meters tracking parameters like $O^{CN\_SoC}_t$, $O^{CN\_EDT}_t$, $O^{CN\_SoC\_D}_t$. Beyond these local insights, agents can also connect to a server/cloud to get real-time $O^{EP}_t$, and $O^{CE}_t$. Edge computing at the dwelling not only ensures reduced latency, crucial for rapid, continuous decisions, but also supports offline functionality and control, if necessary. By processing local sensor data on-site it maintains sovereignty, ensuring that sensitive information remains within the dwelling. Additionally, this decentralized structure diminishes the risk of system-wide vulnerabilities or failures, as issues in one unit do not directly compromise the entire REC.

*C. Multi-Agent Deep Deterministic Policy Gradient*

The proposed MARL EnergAIze approach is anchored on the Multi-Agent Deep Deterministic Policy Gradient (MADDPG) [25] algorithm, which evolved the DDPG single-agent algorithm into a centralized training with decentralized execution. MADDPG is tailored for environments that balance cooperation and competition between decentralized agents, making it an appropriate choice for this work's objectives (cooperating for optimizing REC goals such as balancing and competing for individual local optimization based on prosumer's personal objectives).

*1) Training and Execution Phases*

Fig. 3. evolves from Fig. 1. and presents the central components of the MADDPG's algorithm framed within an energy management context. It has two distinct phases represented: the execution phase (surrounded by a green dashed line) and the training phase (surrounded by a purple dash line). The biggest change from Fig. 1. architecture is in the way agents are represented. In Fig. 3. each agent has two deep neural networks: an Actor Network (μi), used during the execution and training phases, and a Critic Network (Qi), used for training the algorithm. In the MADDPG, the Actor networks decide on the most appropriate actions for their corresponding agents, effectively translating to having one policy function per dwelling at deployment runtime, where personal objectives and prosumer routines are learned by the neural network. The Critic networks guide the learning process, and assess the global consequences of local actor decisions, also ensuring the objective of REC optimization.

*a) Execution Phase*

During the execution (inference) phase the decentralized agent $i$, now represented by the deep neural Actor Network (μi), processes $O^i_t$ by consulting its policy function $\pi^i(A^i_t|O^i_t)$ to infer on the actions to take $A^i_t$. Once decided, $A^i_t$ are applied to the environment, and $R^i_t$ calculated.

Besides the goal of outputting actions for energy management, μi, is also responsible for Experience Collection, a key step into a continued learning and improvement of EnergAIze. At each time step $t$, an experience is gathered and stored in a memory structure called the Replay Buffer. Each recorded experience encompasses $S_t$, which aggregates $O^i_t$ from all the prosumers (agents) within the EC at time $t$, $A_t$, which aggregates the actions $A^i_t$ taken by all agents in the EC at time $t$, $R_t$, which aggregates the rewards $R^i_t$ of all agents in the EC after $A^i_t$ are applied, $S_{t+1}$ which aggregates $O^i_{t+1}$ within the EC at time $t+1$ after $A^i_t$ being applied.

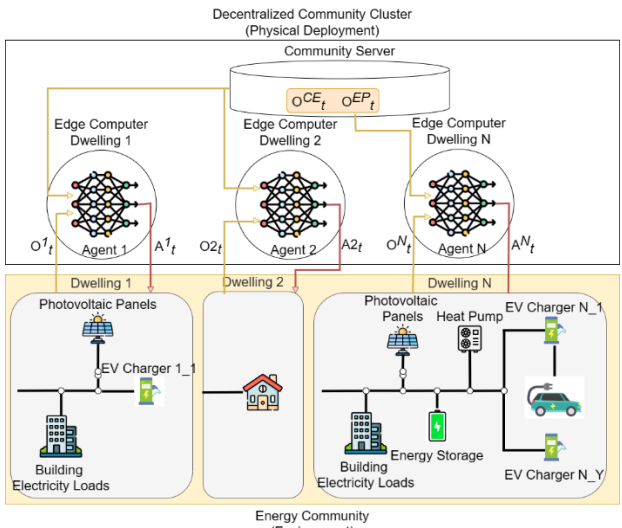

*Figure 2 - Physical deployment of EnergAIze at the edge*

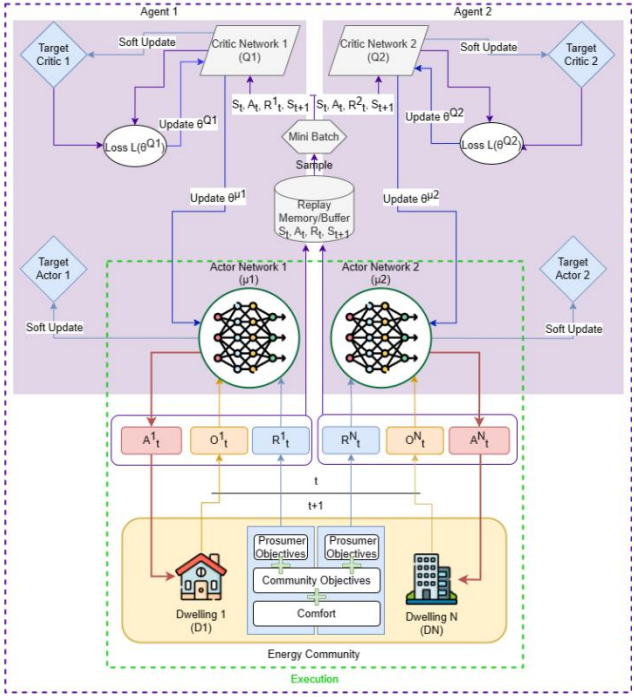

*Figure 3 - Detailed implementation of the MADDPG*

*b) Training Phase*

Once sufficient experiences have been saved in the Replay Buffer during the execution phase (determined by the *batch_size* hyperparameter), the algorithm can utilize these experiences to update (train) the agents' policy. To ensure stable training, the algorithm uses soft updates and target networks [25]. During training three things happen:

1. **Sample Experiences:** First, a mini batch of experiences is sampled from the Replay Buffer for each agent *i*. Each sampled experience consists of four components: individual reward, $R^i_t$, and global $S_t$, $A_t$, and $S_{t+1}$.
2. **Critic:** Following the conventional critic network update process using sampled experiences, the Critic network produces an estimate $Qpred(S_t, A_t)$. Alongside this, using the Bellman equation, a target Q-value, $Qtarget$, is derived considering the upcoming state $S_{t+1}$, individual reward $R^i_t$, and the discount factor. Then, the Temporal Difference (TD) error measures the difference between these values to represent the magnitude of the estimation error. Once squared to form the Loss, it directs the adjustment of the Critic's forecasts.
3. **Actor:** For each agent, the objective of the Actor network is to identify the action that maximizes the expected Q-value as determined by its Critic, through gradient ascent. Specifically, the Actor derives the gradient of the Q-value concerning its actions and then employs this gradient to update its network parameters.

*2) Actor and Critic Networks Architecture*

Fig. 4. depicts the architecture of the networks. Each agent *i* is outfitted with an Actor Network that guides its optimization by processing observations with an input layer of width $O^i_t$. Then, it is followed by a set of fully connected hidden layers using ReLU activation functions for non-linear modeling. The output layer, consisting of neurons equal to the number of possible actions $A^i_t$. The Critic Network evaluates the actions suggested by the actor within a broader scope of the REC. It features a complex architecture with two initial input layers: a State Layer for environmental dynamics (width

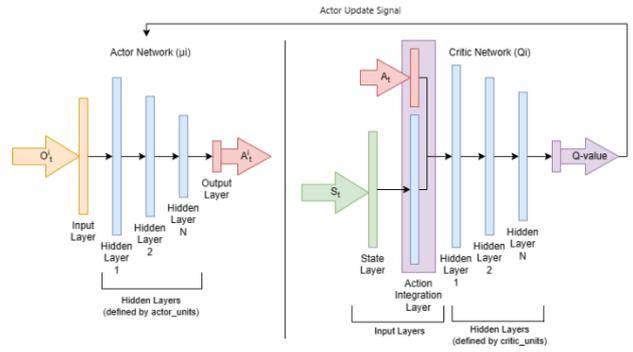

*Figure 4 - Actor and Critic networks architecture for EnergAIze*

of $S_t$, as the total number of observations of the community) emphasizing State Primacy, and an Action Integration Layer to incorporate actions from all agents in the REC at time *t* ($A_t$). The following are fully connected hidden layers (specified by *critic_units* hyperparameter) for in-depth learning. The output is a single Q-value, estimating the cumulative reward for the actions of the specific dwelling.

*3) Hybrid MADDPG-RBC approach for Exploration*

This work adopts a hybrid exploration training (Fig. 5.) strategy that mixes the conventional Gaussian noise, with a Rule Based Controller (RBC) guided exploration in the beginning. By relying on the RBC, presented in [23], the agent can skip the initial phase of random exploration, which can be time-consuming (solutions that are not accepted within the problem constrains, such as leaving without the required SoC). Relying solely on the RBC means that the algorithm will not explore beyond the policy of the RBC, and effectively it will not learn past it. I. The RBC is used in the beginning of exploration. Gaussian noise is applied after to observation values and decays in magnitude along the training [23].

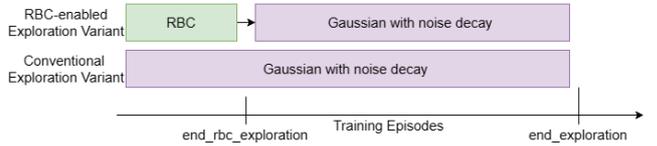

*Figure 5 – RBC enabled and Conventional exploration variants*

D. *Reward Function*

The reward function evaluates all agents' actions to manage individual objectives while contributing to REC goals and steering towards the hard context constraint rules imposed by the complexity of EV charging (e.g., the car cannot charge if it is not present). Eq. (1) encapsulates the three crucial components of EnergAIze's mixed reward function, comprehensively detailed in [23].

$$R^i_t = \alpha \times rProsumer + \beta \times rEV + \zeta \times rREC \quad (1)$$

Here, $R^i_t$ indicates the agent *i* reward for a specific timestep *t*. The *rProsumer* component caters to the unique objectives or performance of each prosumer. Three different components are prepared given the unique choice of the prosumer. Prosumers can choose from i) cost minimization goal, ii) self-consumption maximization goal, and iii) carbon emissions reduction goal. The *rEV* component of the reward embodies constraints associated with EV operations, like meeting charging needs. The *rREC* encapsulates broader REC goals, such as optimizing and balancing grid demand and promoting shared energy resources. The coefficients $\alpha$, $\beta$ and $\zeta$ affect the relative significance of individual versus community objectives and were object of fine tuning in [23].

## IV. RESULTS

### A. Simulation Environment

In this work, a data fusion process was accomplished to construct a Simulation Scenario (SS) compatible with the CityLearn framework [26]. CityLearn is a standardized environment for facilitating benchmarking of RL and MARL algorithms for DR. At its core CityLearn does not simulate EVs, and as such, EVLearn extension is used.

The base SS derives from a dataset previously used within CityLearn [26]). It features 9 dwellings with 4 years of data. The dwellings are of different types, including a medium-sized office, a fast-food restaurant, a standalone retail store, a strip-tease mall, and five medium-scale multifamily residences. The dataset contains date related with different devices, like air-to-water heat pumps, electric heaters, and PVs. For V2G simulation, the dataset includes simulated EVs, EVCs and their energy flexibility. Finally, each prosumer (dwelling) was attributed a different personal objective. Refer to Appendix J of [23] for detail at each prosumer flexible assets and individual objectives. The original dataset was further enriched by integrating Real Time energy pricing from the Iberian wholesale energy market (OMIE)[1].

To evaluate the success of the objectives, a set of Key Performance Indicators (KPIs) have been established according to the KPIs defined originally in [20]. KPIs are presented as a normalized value to the SS baseline. The baseline, in this context, represents the initial data observed and measured, indicating the EC consumption prior to the introduction of any EMS. Three algorithms will be compared: EnergAIze (ENER), MARLISA (MAR) and a Soft-Actor Critic Network (SAC) trained for 15 episodes. Results are normalized to the no control baseline of the SS during the final deterministic episode (i.e., the algorithm simulates the 4 years of data without learning, applies previously learned behaviors and evaluates comparing to the baseline).

### B. Experimental Results With V2G

Fig. 6 outputs the management V2G results for a day of dwelling 7 (D7) within the dataset, with personal cost minimization objectives, using EnergAIze. The green zones are when the vehicle was connected to the dwelling. At the top plot blue squares represent the actual SoC of the EV and the red dots the required SoC at departure. At the bottom plot, the real-time electricity prices for the same day are depicted.

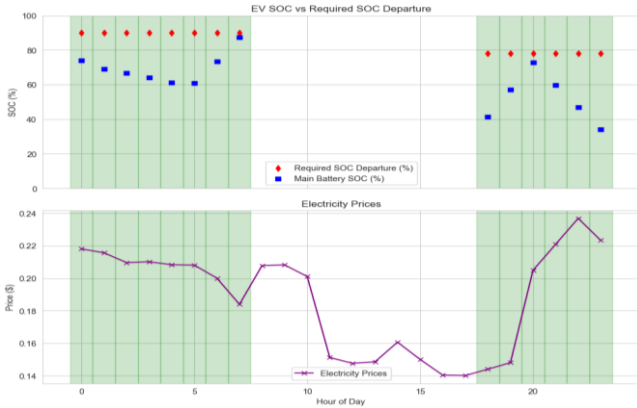

Figure 6 – V2G energy management

---

[1]https://www.omie.es/en/market-results/daily/daily-market

Refer to Fig. 6, at 8:00 am and verify how the EV leaves with a SoC close to what was required by the prosumer. Also refer to Fig. 6 at 6:00 am and from 6:00 pm to 8:00 pm, and check how the algorithm learned to charge the battery at times of lower electricity prices, and the discharging happening around 8:00 pm when prices increase and the dwelling consumption is higher (peak reduction and cost reduction). Table 1 resumes the normalized KPIs at the REC

Table 1 – Comparison of REC-level KPIs

| KPI | SAC | MAR | ENER |
|---|---|---|---|
| Electricity Consumption (D) | -10.30% | -7.35% | -12.46% |
| Electricity Price (C) | -9.39% | -6.13% | -11.35% |
| Carbon Emissions (G) | -10.12% | -7.14% | -11.84% |
| Zero Net Energy(Z) | +4.72% | +2.85% | +6.22% |
| Average Daily Peak (P) | -18.95% | -13.65% | -20.80% |
| Ramping (R) | -29.29% | -25.99% | -35.22% |
| 1 - Load Factor (1-L) | -11.82% | -8.97% | -13.43% |

The total carbon emissions (G) and energy cost were cut by 12% compared to the baseline. Zero net energy (Z) also recorded an enhancement, marked by a 7% decrease against baseline numbers. Of particular interest is the average ramping metric (R), the daily peak average (P) and the 1-Load Factor (1-L), where the EnergAIze exhibited a reduction of 35%, 21% and 13,5% respectively. Moreover, when set side by side with the other algorithms, EnergAIze exhibited superior performance by a margin of 2 to 7%.

Table 2 presents the reduction in percentage of the dwelling level KPIs (from Dwelling 1, D1, to Dwelling 9, D9) at, electricity cost (C), carbon emissions (G), and zero net energy (Z). In the color coding of Table 2, red means it is the worst compared to the other t algorithms in that specific metric for that specific dwelling, while yellow means it is the second best and green illustrates the best result. Below the building identification is the reference to the personal goal of each.

Out of the nine dwellings, the EnergAIze algorithm demonstrated superior performance in most cases. For Dwelling 1 and Dwelling 2, the primary objective was to reduce energy costs (C). The EnergAIze algorithm successfully achieved this objective, registering reductions of 17.3% and 14.1% respectively, and surpassed the performance of other algorithms for that specific KPI. Building 4, categorized as a strip mall, was driven by an environmental goal to minimize its carbon footprint (G). Here too, EnergAIze reduced the building's carbon emissions by 26.7%.

Table 2 - Dwelling-level Results

| | D1 (C) | D2 (C) | D3 (G) | D4 (G) | D5 (C) | D6 (Z) | D7 (C) | D8 (Z) | D9 (Z) |
|---|---|---|---|---|---|---|---|---|---|
| Energy Cost (C) % | | | | | | | | | |
| SAC | -15,60 | -11,64 | -7,26 | -23,60 | -7,47 | -3,87 | -1,91 | -4,60 | -3,94 |
| MAR | -13,57 | -10,49 | -3,04 | -9,79 | -3,68 | -0,67 | -0,37 | -0,88 | -0,97 |
| ENER | **-17,25** | **-14,08** | -6,29 | **-25,69** | **-10,41** | -6,20 | **-4,22** | -7,11 | -6,19 |
| Carbon Emissions (G) % | | | | | | | | | |
| SAC | -15,98 | -11,88 | **-8,72** | -25,01 | -8,36 | -4,29 | -2,34 | -4,96 | -4,52 |
| MAR | -17,78 | -14,04 | -2,80 | -10,81 | -3,79 | -0,79 | -0,35 | -0,75 | -0,85 |
| ENER | -17,09 | -13,77 | -7,94 | **-26,69** | -11,36 | -6,73 | -4,25 | -7,33 | -6,44 |
| Zero Net Energy (Self-Consumption) (Z) % | | | | | | | | | |
| SAC | -5,68 | -4,86 | -4,32 | -9,15 | -4,69 | -3,42 | -2,11 | -4,38 | -3,75 |
| MAR | -6,43 | -6,93 | -0,25 | -0,87 | -1,06 | -0,23 | -0,21 | -0,26 | -0,29 |
| ENER | -6,45 | -6,69 | -3,37 | -9,07 | -7,27 | **-5,90** | -4,19 | **-6,92** | **-5,93** |

In the cases of Dwellings 5 and 7, where the primary objective was also to reduce energy costs (C), the successes mirrored those of Buildings 1 and 2, with EnergAIze realizing significant reductions in energy costs. Buildings 6, 8, and 9, which were oriented towards consuming more self-produced energy, saw their objectives met by the actions of the algorithm. For Building 3, which was aimed at reducing carbon emissions, EnergAIze did not emerge as the top performer in comparison to other algorithms. Yet, it still managed a noteworthy reduction in carbon emissions for this building (7.94% compared to SAC's 8.72%).

## V. Conclusions

This paper presented EnergAIze, a MARL energy management framework. EnergAIze enables individual prosumers to chase their personal goals, yet collaboratively working towards achieving EC objectives. Moreover, the paper presented the planned decentralized architecture deployment. Results provided an exploration of the EnergAIze algorithm's applicability and contributions for V2G scenarios within RECs. EnergAIze. showed diminished peak consumption and ramping while attending specific individual goals of the prosumers.

Moving forward, it is imperative to enrich EnergAIze's development by testing it in a variety of complex environments. This involves incorporating more real-world data, applying the algorithm in actual edge devices of a REC, increasing simulation time step granularity, and testing for learning transferability. Additionally, incorporating battery wear into the reward function, handling interactions with external energy demands, and striving for explainable AI are crucial. These steps are key to proving EnergAIze's effectiveness and its ability to adapt to the real-world.

## Acknowledgment

This paper is supported by the OPEVA project that has received funding within the Chips Joint Undertaking (Chips JU) from the European Union's Horizon Europe Programme and the National Authorities (France, Czechia, Italy, Portugal, Turkey, Switzerland), under grant agreement 101097267. The paper is also supported by Arrowhead PVN, proposal number 101097257. Views and opinions expressed are however those of the author(s) only and do not necessarily reflect those of the European Union or Chips JU. Neither the European Union nor the granting authority can be held responsible for them. The work in this paper is also partially financed by National Funds through the Portuguese funding agency, FCT - Fundação para a Ciência e a Tecnologia, within project LA/P/0063/2020. DOI10.54499/LA/P/0063/2020, https://doi.org/10.54499/LA/P/0063/2020